\documentclass[printer]{aa}
\usepackage{natbib,graphicx}
\begin{document}
   \title{3D outer bulge structure from near infrared star counts}

   \author{S. Picaud\inst{1}\inst{2} \and A. C. Robin\inst{1}}

   \institute{CNRS UMR6091, Observatoire de Besan{\c c}on, BP 1615,
    F-25010 Besan{\c c}on Cedex, France\\
              \email{picaud@obs-besancon.fr,annie.robin@obs-besancon.fr}
              \and
	      Astronomisches Rechen-Institut, M\"onchhofstra{\ss}e 12-14,
D-69120 Heidelberg, Germany\\
              \email{picaud@ari.uni-heidelberg.de}}

   \date{Received 4 May 2004 ; accepted 17 June 2004}

   \abstract{We attempt to study the characteristics of the different
stellar populations present in the Galactic central region.
A Monte Carlo method is used to simultaneously fit
11 thin disc and triaxial outer bulge density parameters on
(K$_s$, J-K$_s$) star count data in almost 100 windows from the
DENIS near infrared large scale survey
at -8$^\circ$$<$l$<$12$^\circ$ and $|$b$|$$<$4$^\circ$.
Various bulge density profiles and luminosity functions were tested
using a population synthesis scheme.
The best models, selected by a maximum likelihood test,
give the following description: the outer bulge is
boxy, prolate, and oriented 10.6$^\circ\pm$3$^\circ$ with respect to
the Sun - center direction. It seems that the main bulge population
is not older than 10 Gyr,
but this preliminary result needs further work to be confirmed.
A significant central hole is found in the middle
of the thin disc.
We discuss these results in regard to previous findings and the scenario
of bulge formation.
   }

\titlerunning{The shape of the Galactic bulge from DENIS}

   \maketitle
%
%________________________________________________________________

\section{Introduction}

Due to the high extinction as well as the superimposition of different stellar
populations in the region, the structure of the inner Milky Way
remains uncertain. In the last 10-15 years, the use of infrared data, less
sensitive to extinction, made it possible to make much progress in our
knowledge of the bulge region. However, some questions such as the precise
orientation and the length of the outer bulge, as well as the existence
and the length of the central disc hole have received contradictory
answers.

There is now a consensus that the outer bulge is triaxial:
Blitz \& Spergel (1991) and Kent et al. (1991), using 2.4 $\mu$m maps, and
Nakada et al. (1991), using IRAS stars, detected asymmetries in longitude
which they explained by the triaxiality of the outer bulge,
also called the bar, with the near end at positive longitudes ;
Binney et al. (1991)
studied the gas kinematics and came to the same conclusion.
Then, studies using COBE/DIRBE maps (e.g. Dwek et al. 1995;
Binney et al. 1997; Freudenreich 1998; L\'epine \& Leroy 2000;
Bissantz \& Gerhard 2002), star counts (e.g. Nikolaev \& Weinberg 1997;
Stanek et al. 1997; L\'opez-Corredoira et al. 2000),
kinematics (e.g. Deguchi et al. 2002; Zhao 1996)
and microlensing (e.g. Zhao \& Mao 1996) made it possible
to deduce a more detailed description of the triaxial bulge/bar,
and in particular gave estimations
of its orientation in the Galactic plane. But, if all studies converge to
the description by a prolate shaped outer bulge with the major axis almost
lying in the Galactic plane and the near end in the first quadrant,
the estimated values of the angle of the major axis
from the Sun direction vary between about 10$^\circ$ and 30$^\circ$.
Some studies even present a bar with an angle around 40$^\circ$-45$^\circ$
(e.g. Nakai 1992; Weinberg 1992; Deguchi et al. 1998;
Sevenster et al. 1999; L\'opez-Corredoira et al 2001).
The reason for this difference is that they deal with two
different things: L\'opez-Corredoira et al. (2001), then Picaud et al. (2003),
showed the existence of a structure at l=20$^\circ$-27$^\circ$,
which may be at the top end of a long in-plane bar with an angle
of about 40$^\circ$ from the Sun-center direction.
This structure is distinct from the
triaxial outer bulge studied in the present work,
but the confusion often appears in the literature.

Alard (2001) showed evidence for a nuclear bar in the most central
parts of the Milky Way which may correspond to the inner bulge.
But according to Ibata \& Gilmore (1995),
the inner bulge and the outer bulge may be two distinct populations.
Therefore, in order to avoid any confusion with another bar or with
the inner bulge, we will give the name \emph{outer bulge} to the
triaxial prolate structure observed in the inner region ($|$l$|\leq$10$^\circ$)
excluding the very central parts ($|$l$|\leq$1$^\circ$ and
$|$b$|\leq$1$^\circ$).

Another important stellar population present in the outer bulge region is
the thin disc. The shape of the inner thin disc is still a controversial issue.
In particular, the existence of a truncation or a hole at its
center is still in doubt, even if there are more and more indications
in this direction, in particular in external galaxies.
For instance, Freeman (1970) classified the discs in two types, the first
described by a single exponential, and the second modeled by
the subtraction of two exponentials, i.e. with a central hole.
Ohta et al. (1990), studying 6 early-type spiral
galaxies, showed that all had Freeman type II discs.
According to Bagget et al. (1996), the proportion of galaxies having an
inner truncated disc is at least twice as numerous in barred spirals
as in non-barred ones. In our Galaxy, Freudenreich (1998)
fitted his bulge and disc
model on the COBE/DIRBE map and found a disc hole radius of 3 kpc.
More recently, L\'epine \& Leroy (2000) showed that a disc model including the
central hole was more convenient to fit
the brigthness distribution and the rotation curve, and
L\'opez-Corredoira et al. (2001) used a disc
model with a truncation inside the Galactic ring at 3.7 kpc.

In this paper, we have compared simulated star counts from the
Besan\c{c}on model of the Galaxy with DENIS near infrared
data in almost one hundred
windows of low extinction distributed between -12$^\circ$ and +8$^\circ$ in
longitude and -4$^\circ$ and +4$^\circ$ in latitude.
Star counts are a better means than surface brightnesses to
determine the 3-dimensional density parameters:
integrated fluxes are dominated by the brightest and closest stars
while star count studies take into account a wider range of intrinsic stellar
luminosities and distances.

The article is organized as follows. In Sect. 2, we
will describe the data (DENIS batches), the selection of low extinction
windows and the selections made in color and magnitude.
In Sect. 3, after a brief description of
the Besan\c{c}on model of the Galaxy, we will present the disc and
outer bulge density distributions and luminosity functions (hereafter LF)
used to compute the simulations and fit the parameters.
In Sect. 4, we will describe the fitting method, based on Monte Carlo
drawings and a maximum likelihood test. Results are given in Sect. 5
and compared with previous studies in Sect. 6. We conclude in Sect. 7.

%__________________________________________________________________

\section{The Data}

\subsection{DENIS Batches}

The DENIS (Deep Near Infrared Survey of Southern Sky) survey
(Epchtein et al. 1997, Fouqu\'e et al. 2000)
covers almost all the Southern Sky (97 \% at the
end of the survey) in strips of 30$^\circ\times$12$\arcmin$. Photometric bands
used are Gunn-I (0.85 $\mu$m), J (1.25 $\mu$m) and K$_s$
(2.15 $\mu$m).

In addition to the survey strips, specific observations called
``batches'' (Simon et al, in preparation)
were made in 1998 in outer bulge and plane regions
in smaller fields (about 2 deg$^2$).
Data reductions were made at PDAC (Paris Observatory Data Analysis Center).
A PSF fitting optimised
for the crowded fields was used for the source extraction. All standard
stars observed in a given night were taken for the determination of the
photometric zero point, resulting in accuracies ranging for 0.08 at
K$_s$=8 to 0.15 at K$_s$=13, and 0.08 at J=10 to 0.15 at J=15.

\subsection{Windows of low extinction}

   \begin{figure*}
   \centering
   \caption{Windows of low extinction (small filled boxes) and batches
(long boxes).
Grey levels of windows represent the A$_V$ given by Schultheis et al. (1999).}
   \includegraphics[angle=-90,width=13cm]{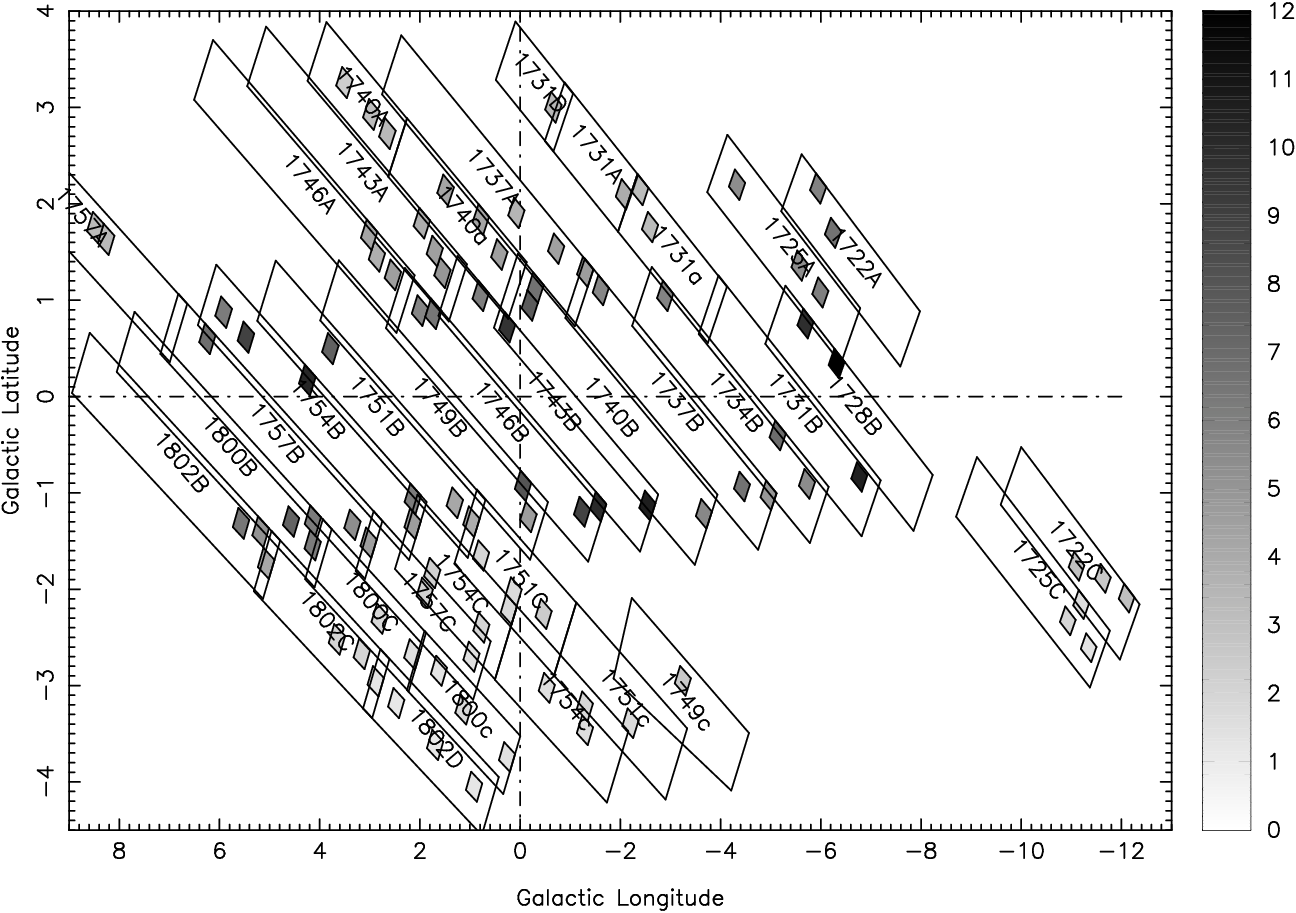}
   \label{fenetres}
   \end{figure*}

94 windows of about 15'x15' between the
longitudes [-12$^\circ$; 8$^\circ$] and the latitudes [-4$^\circ$; 4$^\circ$]
were selected in DENIS batches. These windows were selected from the
Schultheis et al. (1999) extinction map as having either a low or homogeneous
extinction which is easy to model.
For most windows, the extinction distribution along the line of sight
was modeled using no diffuse extinction
but 2 clouds (localized extinction), one at 1 kpc (Sagittarius-Carina arm)
and the other at about 4 kpc (Scutum-Centaurus arm). In a few
windows, diffuse extinction was added.
In each cloud, the $A_V$ was estimated by comparing the quantiles of
J-K$_s$ color between data and simulations (see Sect. \ref{modele}).
The reddening in each photometric band was taken
from the extinction law of Mathis (1990).

Unfortunately, as one can see in Fig. \ref{fenetres},
there are almost no windows selected very close to
the Galactic plane ($|$b$|<$1$^\circ$), because of the large extinction
there.

\subsection{Photometric bands and star selections}

In the present study, only magnitude K$_s$ and color J-K$_s$ were used to
compare observed star counts with simulated ones, the I band being too
sensitive to extinction.

The blue side of the color-magnitude diagram (hereafter CMD) is mostly
populated by foreground dwarfs, especially at faint magnitudes.
Cuts were made in K$_s$ and J-K$_s$ to reduce this contamination:
we kept only stars between 7.5 and an upper limit varying from field to field
(12.5 on average) in K$_s$, and with
J-K$_s$ $>$ J-${\mbox{K}_s}_{/lim}$, J-${\mbox{K}_s}_{/lim}$ being
in the mean equal to 0.8 but also varying from
field to field due to extinction.
To ensure completeness, only stars below $\approx$15 mag in J have been kept.

%__________________________________________________________________

\section{The model}\label{modele}

The Besan\c{c}on model of the Galaxy has been used to compute simulations
and compare them with the data. It is described in Sect. \ref{besac}.
The holed thin disc and triaxial outer bulge density models which have
been fitted are presented there, as well as the luminosity function used.

\subsection{The Besan\c{c}on model of the Galaxy}\label{besac}

The Besan\c{c}on model of stellar population synthesis aims at giving a global
3-dimensional description of the Milky Way structure and evolution,
including stellar populations
such as thin disc, outer bulge, thick disc and spheroid, as well as dark halo
and interstellar matter. 
Details can be found in Robin \& Cr\'ez\'e (1986), Bienaym\'e et al. (1987)
and Robin et al. (2003,2004).

The approach of the Besan\c{c}on model is semi-empirical: both theoretical
schemes (stellar evolution, galactic evolution, galactic dynamics) and
empirical constrained are used.
Boltzmann and Poisson equations make it possible to self-consistently constrain
the disc scale height via the Galactic potential.
Simulated catalogues of stars are built and give
observables (apparent magnitudes, colors, proper motion,
radial velocities) directly comparable with the data. Interstellar extinction,
photometric errors and Poisson noise are also added to make simulations 
as close as possible to observations.

The Besan\c{c}on model has been used to determine the density laws and
luminosity functions of the stellar populations: thin disc
(Robin, Cr\'ez\'e \& Mohan 1992,
Ruphy et al. 1996, Haywood et al. 1997), thick disc
(Robin et al. 1996, Reyl\'e \& Robin 2001),
spheroid (Robin et al. 2000), and outer bulge in the present work.

\subsection{Holed Thin Disc}

The thin disc may contribute significantly to star counts in the Galactic
central region. Hence its characteristics have to be fitted at the same time
as the bulge parameters.
The two disc shape parameters that have an effect on star counts of the
inner region are the scale lengths of the disc ($R_d$)
and of its central hole ($R_h$). Other
parameters and structures such as outer radius, warp and flare are
taken into account in the Besan\c{c}on model but are not relevant here.

\subsubsection{Density distribution}

	The thin disc is divided into 7 age components: the first
(age $<$ 0.15 Gyr) is called \emph{young disc}, the other six forming the
\emph{old disc}\ (0.15 Gyr $<$ age $<$ 10 Gyr). The young thin disc is not
studied here, because
its density is very low in comparison with the bulge and the old disc,
and is probably very patchy.
Fitting its parameters would be difficult and inefficient.

	The old thin disc density distribution model follows the
Einasto (1979) law: the distribution of each old disc component is
described by an axisymmetric ellipsoid with an axis ratio depending on
the age; the density law of the ellipsoid is
described by the subtraction of 2 modified exponentials:
$$\rho_d = \rho_{d_0} \times [\exp (-\sqrt{0.25+
(\frac{a}{R_d})^2})
-\exp (-\sqrt{0.25+(\frac{a}{R_h})^2})]$$
\begin{center}
with $a^2=R^2+\left(\frac{Z}{\epsilon}\right)^2$, where:
\end{center}
\begin{itemize}
\item $R$ and $Z$ are the cylindrical galactocentric coordinates.
\item	$\epsilon$ is the axis ratio of the ellipsoid. Table \ref{tabledisque}
gives the recently revised (Robin et al. 2003)
axis ratios of the 6 age components of the old thin disc.
\item	$R_d$ is the scale length of the disc and is around 2.3-2.5 kpc
(Ruphy et al. 1996).
\item	$R_h$ is the scale length of the hole.
\item The normalization $\rho_{d_0}$ is deduced from the local luminosity
function (Jahrei{\ss} et al, private communication), assuming that the Sun is
located at R$_\odot$=8.5 kpc and Z$_\odot$=15 pc. Local densities are
given in Table \ref{tabledisque}.
\end{itemize}

\begin{table}[!h]
\caption{Axis ratios and local densities of the six age components of the
old thin disc.}
{\centering \begin{tabular}{ccc}
\hline
\hline
\textbf{age (Gyr)}& $\epsilon$ & $\rho_{d_0}$ ($\star$.pc$^{-3}$)\\
\hline
0.15-1 	& 0.0268 & 0.03146 \\
1-2	& 0.0375 & 0.02538 \\
2-3	& 0.0551 & 0.01887 \\
3-5	& 0.0696 & 0.02625 \\
5-7	& 0.0785 & 0.02037 \\
7-10	& 0.0791 & 0.02284 \\
\hline
\label{tabledisque}
\end{tabular}\par}
\end{table}

\subsubsection{Luminosity Function}

\begin{figure}[h!]
\centering
\caption{Luminosity functions of the thin disc
in the K$_s$ band. The young disc
corresponds to the dashed line, while the old one
is represented by the dotted line.
The solid line represents the total thin disc LF.
On the ordinate: the decimal logarithms of
the numbers of stars per 1 mag bin of absolute magnitude.}
\includegraphics[angle=-90,width=8cm]{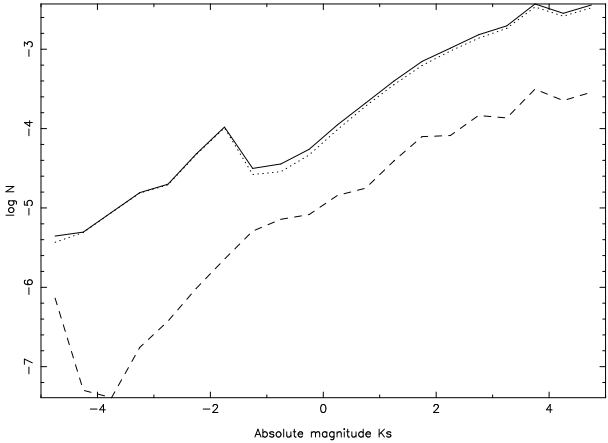}
\label{FL-disque}
\end{figure}

A standard evolution model is used to produce the disc population,
based on a set of evolutionary tracks,
a constant Star Formation Rate (hereafter SFR)
and a two-slope Initial Mass Function (hereafter IMF)
$\phi(M)=A\cdot M^{-\alpha}$ with $\alpha$=1.6 for M$<$1$\cdot M_\odot$ and
$\alpha$=3.0 for M$>$1$M_\odot$.
The preliminary tuning of the disc evolution parameters against relevant 
observational data was described in Haywood et al. (1997)
and further changes are explained in Robin et al. (2003).

Magnitudes and colors in various filters are computed
using semi-empirical model atmospheres from
Lejeune et al. (1997, 1998).

Fig. \ref{FL-disque} presents the luminosity functions in
K$_s$ of the young disc, the old disc and the thin disc in totality.
This figure confirms how dominant the contribution of the old component
on the thin disc is,
in particular at the magnitudes [-5;-2] which are the most frequent absolute
magnitudes in the simulations of the present study.
Within the range of apparent
magnitude and color used in the present study, observed stars have absolute
K$_s$ magnitudes brighter than -1.

\subsection{Triaxial bulge}\label{bulbe}

To simulate the star counts, the bulge density law and a
luminosity function have to be assumed. The different bulge density profiles
used are presented in Sect. \ref{densite}. 5 different luminosity
functions have been tested, as explained in Sect. \ref{bulbeFL}.

\subsubsection{Density distribution}\label{densite}

\medskip
\textbf{Orientation}
\medskip

There is consensus that the bulge is triaxial.
Three angles define the orientation:
\begin{itemize}
\item	$\phi$: orientation angle
from the sun-center direction of the projection
on the Galactic plane of the bulge major axis.
\item	$\beta$: pitch angle of the bulge major axis from the Galactic
plane. In all previous studies, $\beta$ was found to be
very close to 0$^\circ$.
\item   $\gamma$: roll angle around the bulge major axis
\end{itemize}

However, the third angle $\gamma$ is ill defined because
the minor axes have similar scale lengths.
Hence we prefered to fix it at $\gamma$=0$^\circ$, and only $\phi$ and
$\beta$ have been fitted.

   \begin{figure*}
   \centering
   \caption{Definition of the angles to pass from the Galactic frame
(x,y,z) to the bulge frame (X,Y,Z). The
transformation consists of 2 consecutive rotations:
the first is a clockwise
rotation of $\phi$ around the galactic vertical axis z,
and the second is a clockwise rotation of $\beta$ around the new axis y'.}
   \includegraphics[width=10cm]{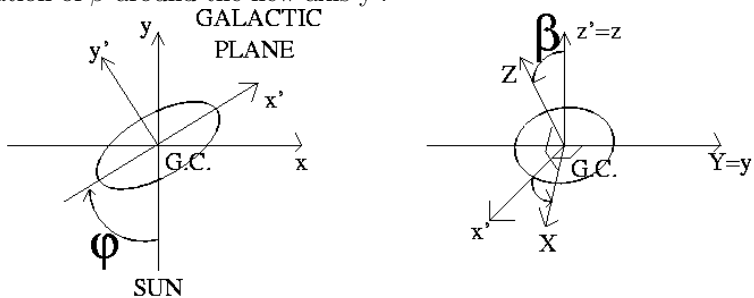}
   \label{angles}
   \end{figure*}

\medskip
\textbf{Density profiles}
\medskip

Dwek et al. (1995) and Freudenreich (1998)
fitted various density distributions to the
near infrared surface brightness observed using the Diffuse Infrared Background
Experiment (DIRBE) of the Cosmic Background Explorer (COBE).
The best-fitting models obtained at 2.2$\mu$m were
the $G_2$ and $E_1$ functions by Dwek et al. (1995)
and the $S$ function in Freudenreich (1998).
Stanek et al. (1997) tested the same functions of Dwek et al. (1995) using
star counts
of red clump giants. Their best-fitting model, consistent with the observed
star counts as well as the surface brightnesses used by Dwek et al. (1995),
was the $E_2$ function.

We choose to test the 3 different functions
used in these models, as presented in Table \ref{profils}:
an exponential function (called $E$) like $E_1$
and $E_2$, a gaussian one (called $G$) like $G_2$,
and the $S$ function of Freudenreich (1998), which is a sech$^2$
function.

\begin{table*}
\begin{center}
\caption{The 3 bulge density profiles used}
\label{profils}
\begin{tabular}{c}
\hline
\hline
$
\left.
\begin{array}{cc}
E & \rho_E = \rho_0 \times \exp (-R_s)\\
G & \rho_G = \rho_0 \times \exp (-0.5\cdot R_s^2)\\
S & \rho_S = \rho_0 \times \mbox{sech}^2 (-R_s)
\end{array}
\right\}
\mbox{with } R_s^{C_\parallel}=[|\frac{X}{x_0}|^{C_\perp}+
|\frac{Y}{y_0}|^{C_\perp}]^{C_\parallel/C_\perp}+
|\frac{Z}{z_0}|^{C_\parallel}
$\\
\hline
\end{tabular}
\end{center}
\end{table*}

All the best density profiles obtained by Dwek et al. (1995), Stanek
et al. (1997) and Freudenreich (1998) are included in the 3 functions
described above: $E_1$ corresponds to the
$E$ function with $C_\parallel=C_\perp=1$ (diamond shape),
$E_2$ is obtained using the $E$ function and
$C_\parallel$=$C_\perp$=2 (ellipsoidal shape),
and the boxy profile $G_2$ using the $G$ function and
$C_\parallel$=4 and $C_\perp$=2.
The best values obtained in Freudenreich (1998), using the
S function, are:
$C_\parallel$=3.501$\pm$0.0016 and $C_\perp$=1.574$\pm$0.0014.

\

The density function is then multiplied by the cut-off function
$f_c$ (distances are given in kpc, and $R_c$ is called the cut-off radius):

\begin{center}
\begin{tabular}{rl}
$\rho=\rho \times f_c(R_{XY})$, & \textbf{with} $R_{XY}=\sqrt{X^2+Y^2}$\\
$R_{XY} \leq R_c$ $\Longrightarrow$ & $f_c(R_{XY})=1$\\
$R_{XY} \geq R_c$ $\Longrightarrow$ &
$f_c(R_{XY})=\exp \left(-(\frac{R_{XY}-R_c}{0.5})^2\right)$\\
\end{tabular}
\end{center}

\medskip

Finally, the two angles $\phi$ and $\beta$, the three scale lengths
$x_0$, $y_0$, $z_0$,
the density at the center $\rho_0$, the cut-off radius $R_c$ and
the two coefficients $C_\parallel$ and $C_\perp$ are considered
as free density parameters in the fitting process.

\subsubsection{Luminosity functions}\label{bulbeFL}

Star counts are a function of both the density law and the luminosity
functions.
	Five theoretical bulge luminosity functions have been tested, all
of them based on a Salpeter IMF ($\alpha$=2.35) and
assuming a single epoch of formation (starburst) as well as
a mean solar metallicity (Z$\approx$0.02).
Only stellar evolutionary tracks and bulge age vary from one LF to another:

\begin{itemize}
\item Three of them (Fig. \ref{Pad}) have been deduced from theoretical
isochrones by the Padova team (Girardi et al. 2002), and three bulge ages were
tested: 7.9 Gyr, 10 Gyr and 12.6 Gyr, which we shall name Pad7.8, Pad10
and Pad12.6 respectively. These luminosity functions were computed
from the evolutionary tracks of Girardi et al. (2000) for the mass range
$0.15 M_\odot-7.0 M_\odot$ and Bressan et al. (1993) for $M>7.0 M_\odot$,
deduced from a combination of the Girardi et al. (2000)
(with Z=0.019 and Y=0.273), and the Bertelli et al. (1994)
(Z=0.02, Y=0.280) non-$\alpha$-enhanced isochrones.
Stellar atmosphere models
were taken from the ATLAS9 theoretical spectra library (Castelli et al. 1997).

\begin{figure*}
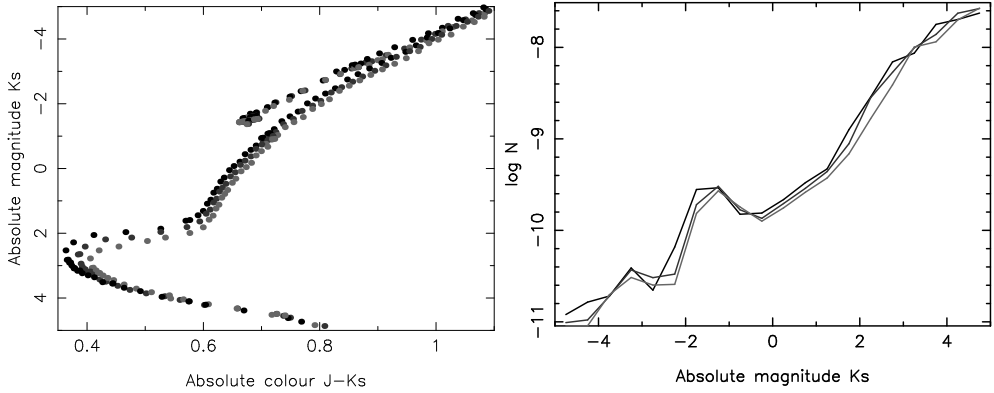

\centering
\caption{\small
Color-magnitude diagrams (left) and luminosity functions (right)
from Padova models.
The three models, Pad7.9, Pad10 and Pad12.6, are shown
in the graphs using dark grey, grey and light gray respectively.
The luminosity functions are given in number of stars per 0.5 mag bin of
absolute magnitude K$_s$.}
\vspace{0.5cm}
\begin{minipage}{6.5cm}
\includegraphics[angle=-90,width=6.5cm]{1218fig4.ps}
\end{minipage}
\begin{minipage}{6.5cm}
\includegraphics[angle=-90,width=6.5cm]{1218fig5.ps}
\end{minipage}
\label{Pad}
\end{figure*}

\

\item The others two (Fig. \ref{Bru}) have been taken from
evolutionary bulge synthesis models of Bruzual \& Charlot
(see Bruzual et al. 1997).
Models were constrained on several bulge globular clusters.
Two bulge ages were tested: 10 Gyr and 12 Gyr, respectively
called BC10 and BC12. Evolutionary tracks were deduced from the Padova 1994
set (Alongi et al. 1993; Bressan et al. 1993; Fagotto et al. 1994a,b;
Girardi et al. 1996), with Z=0.02 and Y=0.280. Stellar atmosphere models
were taken from the semi-empirical Lejeune et al. (1997,1998)
spectral library.

\begin{figure*}
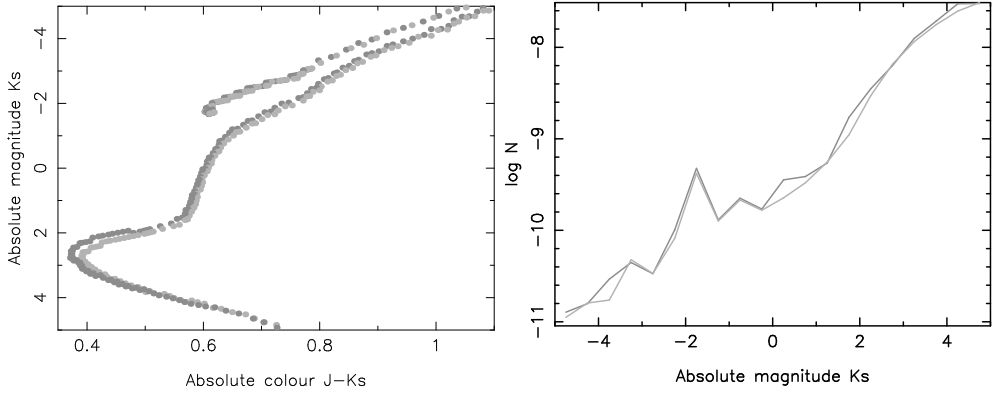

\centering
\caption{\small
Color-magnitude diagrams (left) and luminosity functions (right) from
the Bruzual \& Charlot models. The darker grey corresponds to BC10
and the lighter grey to BC12.
The luminosity functions are given in number of stars per 0.5 mag bin of
absolute magnitude K$_s$.}
\vspace{0.5cm}
\begin{minipage}{6.5cm}
\includegraphics[angle=-90,width=6.5cm]{1218fig6.ps}
\end{minipage}
\begin{minipage}{6.5cm}
\includegraphics[angle=-90,width=6.5cm]{1218fig7.ps}
\end{minipage}
\label{Bru}
\end{figure*}
\end{itemize}

While Bruzual \& Charlot luminosity functions are deduced from the
Padova 1994 set of evolutionary tracks, those of Girardi et al. (2002) use
the Padova 2000 models, which are a new version of the Padova 1994 isochrones.
The model atmospheres also differ: Girardi et al. (2002) use the
theoretical spectra from Castelli et al. (1997), while Bruzual \& Charlot
prefer the semi-empirical library from Lejeune et al. (1997,1998).

\begin{figure*}
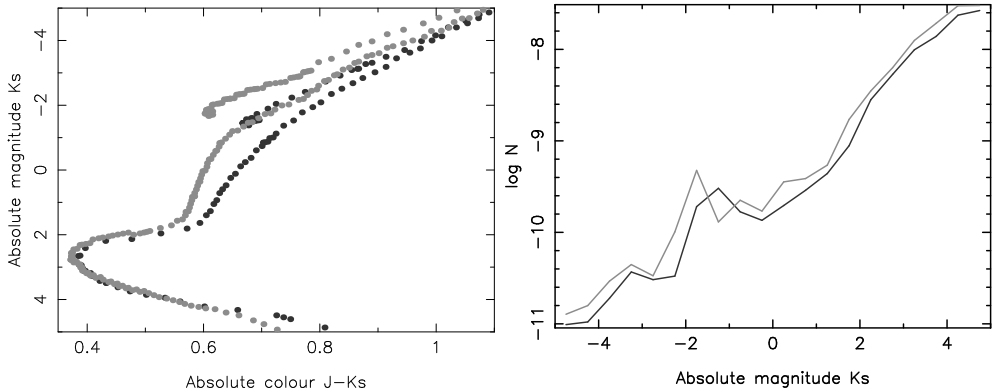

\centering
\caption{\small
Color-magnitude diagrams (left) and luminosity functions
(right) from Padova (Pad10, dark grey) and Bruzual \& Charlot
(BC10, light grey) for an age of 10 Gyr.
The luminosity functions are given
in number of stars per 0.5 mag bin of absolute
magnitude K$_s$.}
\label{FLK-2}
\vspace{0.5cm}
\begin{minipage}{6.5cm}
\includegraphics[angle=-90,width=6.5cm]{1218fig8.ps}
\end{minipage}
\begin{minipage}{6.5cm}
\includegraphics[angle=-90,width=6.5cm]{1218fig9.ps}
\end{minipage}
\end{figure*}

Fig. \ref{FLK-2} compares Bruzual \& Charlot and Girardi et al.
(2002) (Padova) models for an age of 10 Gyr.
In Fig. \ref{Pad}, \ref{Bru} and \ref{FLK-2},
only dwarfs, subgiants, red giants, horizontal branch stars and
AGB stars are shown. Planetary nebul{\ae} and white dwarfs,
included in Bruzual \& Charlot models, are unobservable or negligible in star
counts in the range of apparent magnitude used in the present work. However,
they will be taken into account in the calculation of the total number
of bulge stars in Sect. \ref{totaleffectif}.

The LFs differ only slightly while CMDs are sensitive to the assumed age.
The main age effect is the position of the turn-off,
which corresponds to a color shift of about 0.05 mag for a change
in age of 2 Gyr. The only non-negligible difference between
Padova and Bruzual \& Charlot CMDs is that
the Padova asymptotic giant branch coincides with
the Bruzual red giant branch.

%__________________________________________________________________

\section{Fitting method}

The method used
to fit old thin disc and triaxial bulge parameters to DENIS data
is based on two important points:
firstly, comparisons between data and models are
made with star counts in K$_s$ magnitude and J-K$_s$ color;
secondly, parameters are deduced from a fitting method using Monte Carlo
drawings and maximum likelihood tests.

\subsection{Cuts and choice of bin size}

For each field, selected stars are distributed in 8x2 equally populated
bins: 8 bins of magnitude, and 2 bins of color.
We noticed that the noise relative to the least populated bins increases and
contributes too much to the global likelihood.
This bias increases with the difference between data and model.
To limit this bias,
we group some close windows (always from the same batch) so that we have
at least 70 stars in each bin of magnitude-color.
Approximately 10\% of the windows are grouped together by 2 or 3, and we obtain
88 groups or single windows at the end, for 94 windows in total.

\subsection{Monte Carlo drawings}

    There are 11 parameters to fit:
\begin{itemize}
\item	Bulge orientation: $\phi$, $\beta$
\item	Bulge scale lengths: $x_0$ (major axis), $y_0$ and $z_0$
\item	Bulge normalization $\rho_0$ and cut-off radius $R_c$
\item   $C_\parallel$ and $C_\perp$
\item	Disc scale length $R_d$ and hole scale length $R_h$
\end{itemize}

	An iterative scanning method to explore this 11-dimensional space of
parameters would be too time consuming.
An alternative to save computer
time would be to distribute the parameters in groups
of 3 or 4 and fit them group by group. This method is still too slow
and does not
take into account correlations between parameters, and therefore can
miss some solutions as well as preventing convergence in some cases.
Hence we prefer to use another fitting method based on
Monte Carlo drawings. This method was developed to solve non-linear equations
of star kinematics (Oblak, 1983) and has been adapted for the present study.

	In the Monte Carlo method, parameters are drawn in the 11-dimensional
parameter space.

\begin{itemize}
\item At the first iteration, p points of the 11-D space (or
sets of parameters) are drawn using uniform drawings. Ranges for uniform
drawings are given in Table \ref{intervalles}.

\item At each next iteration, the m points having the highest values of
likelihood among all the points drawn since the beginning are extracted. Then
p new points are determined using semi-gaussian drawings around the
median of the m best points and along the axes defined by the eigenvectors of
their correlation matrix.
Formul{\ae} of median and semi-dispersions related to the semi-gaussian
drawings are given in Appendix \ref{tiragesgaussiens}.

It can happen that drawn values go beyond restrictive limits. In this
case, a new value is determined by a uniform drawing between the
inferior/superior limit and the minimal/maximal value respectively
reached by the m best points on the concerned
parameter. More details about the limits are given in Sect. \ref{poilaunez}.

\item The fit ends when there is no further progress in the likelihood
convergence.
A maximum of 20 iterations is admitted, but this limit is almost never
reached.
\end{itemize}

   Various values of m and p have been tested and the values giving the best
compromise between quality and rapidity of convergence were m=40 and p=300.

\subsubsection{Constraints on parameters}\label{poilaunez}

	Table \ref{intervalles} gives the ranges for the first iteration
uniform drawings.
These ranges were chosen to avoid unrealistic configurations, based on
previous studies which agree on the following description: a
prolate shaped bulge (small $y_0$ and $z_0$ with respect to $x_0$)
with the major axis closer to the Galactic plane (small $\beta$) and its top
end in the first
quadrant (0$^\circ \le \phi \le$ 90$^\circ$).
The following three points, concerning the limits, should be noted:
\begin{itemize}
\item In the case of the angle $\beta$,
the limits apply during the first iteration only.
Subsequent iterations can go beyond it.
\item For certain parameters, such as the central density $\rho_0$
and the bulge scale lengths $x_0$, $y_0$ and $z_0$,
only the inferior limit cannot be passed, and is equal to 0$^+$.
\item In the other cases (the parameters C$_\parallel$ and C$_\perp$,
the angle $\phi$, the cut-off radius R$_c$ and the disc scale lengths
R$_d$ and R$_h$), the limits are kept for all iterations.
\end{itemize}

\begin{table*}
\caption{Ranges for uniform drawings
for the first iteration. Lower limits are given in the first
line while upper limits are given in the second, and restrictions
related to the limits applied in following iterations
in the third. $\infty$ means \emph{no restriction}, and
$[]$ corresponds to \emph{strict limits}.}
\begin{center}
\begin{tabular}{cccccccccccc}
\hline
\hline
 & $\phi$ & $\beta$ & $x_0$ & $y_0$ & $z_0$ &
$\rho_0$ & $R_c$ & $R_d$ & $R_h$ & C$_\parallel$ & C$_\perp$ \\
units & $^\circ$ & $^\circ$ & kpc & kpc & kpc &
$\star$.pc$^{-3}$ & kpc & kpc & kpc & & \\
\hline
inf. & 0 & -10 & 0$^+$ & 0$^+$ & 0$^+$ & 0$^+$ & 1 & 2.2 & 0$^+$
& 1 & 1 \\
sup. & 90 & 10 & 3 & 1 & 1 & 25 & 5 & 3 & 2.0 & 5 & 5 \\
\hline
restrictions & [] & $\infty$ & $>$0 & $>$0 & $>$0 & $>$0 & [] & [] & []
& [] & [] \\
\hline
\end{tabular}
\end{center}
\label{intervalles}
\end{table*}

\subsubsection{Maximum likelihood test}

	Maximum likelihood methods are better than minimum $\chi^2$ ones,
especially in the case of small counts per bin, because minimum $\chi^2$
methods assume that the distribution of observed star counts is a gaussian,
and this hypothesis is false in such a case. Hence,
we preferred to use the maximum likehood to determine the best
sets of parameters in the fitting method,
even if normalized residuals and $\chi^2$ were also used as well as likelihood
to evaluate, after the fits, the agreement between models and observations.

The use of the maximum likelihood method only assumes
that the likelihood is smooth enough and well defined around its local maxima,
which is eventually the case.

$\chi^2$ and likelihood formul{\ae} take into account
the specificities of the study, especially the presence of noise in the
simulations. They are presented in Appendix \ref{vrais}.

\subsection{Weighting}\label{ponderations}

	Computing simulations for each set of parameters drawn
and for all windows would be very time consuming.
Rather, we start with initial simulations, then weightings of these
simulations are applied to adapt them to different density laws.
Initial simulations were done using the G$_2$ function (C$_\parallel$=4,
C$_\perp$=2, $\gamma$=0$^\circ$) from Dwek et al (1995) as the
bulge density distribution, densities with $E$, $G$ and $S$ distributions
being calculated by changing the weightings adequately.

This method may lead to a bias if the
values of parameters used for the initial simulations are
badly chosen: if for instance the initial configuration implies that
there are no bulge stars in a field, then there will be no bulge contribution
for star counts in this field even for a drawn set of parameters which
forecasts the presence of the bulge in the field.
This bias is avoided by making
simulations with appropriate values of parameters, i.e.
no disc hole ($R_h$=0 kpc), bulge scale-lengths all equal to the upper limit
of the major scale-length $x_0$ (3 kpc), large bulge cut-off radius (5 kpc).

A Poisson noise is added to the Besan\c{c}on simulations to make them
resemble the data as closely as possible. Initial simulations were computed
with a density about 5 times greater to increase their signal to noise ratio.

\subsection{Convergence}\label{convergence}

\subsubsection{Dispersions of best parameters}

	Due to the Monte Carlo drawings, the m best sets of parameters of
one fit are not exactly equal but
their values of likelihood are very close, or even the
same. Furthermore, two different fits do not give the same mean of best
values.
That is why 20 independent fits were made, the final result being the
means and dispersions around the 20$\times$m best values.

\subsubsection{Quality of convergence}\label{qualite}

A convergence test was made using simulations to estimate the
correlations and quality of convergence and to identify possible biases in
the procedure.
Three simulations were produced to be used as data and the full
procedure were applied to them to attempt to retrieve the
parameters used for these simulations. The three sets of parameters used
to make these simulations have been chosen from randomly drawn sets
to give three sufficiently different
configurations for the thin disc and the outer bulge.

The main conclusions given by these three tests of convergence and correlations
are the following:
\begin{itemize}
\item The values of the parameters are generally retrieved with good
precision, around 6\% of the range of the initial drawings
and 6\% of the value , except
for some parameters mentioned in the following lines.
\item  The accuracy is good: of the 33 values of parameters
to be found, a third are retrieved at less than 0.5 sigmas,
about 60\% at less than 1 sigma, about 80\% at less than 1.5,
88\% at $\leq$2 and 97\% at $\leq$2.5.
\item Disc scale lengths $R_d$ and $R_h$ are strongly anticorrelated,
which will be taken into account in the discussion.
The disc parameters do not show any
significant correlation with the bulge ones.
However, $R_d$ and $R_h$ best values are sometimes a little biased
with respect to the real values (varying from 0 to 2.4 $\sigma$
difference).
\item $C_\parallel$ and $C_\perp$ are slightly anti-correlated to
$y_0$ and $z_0$ respectively. Moreover, $C_\parallel$ and
$C_\perp$ are not precisely determined, having a dispersion between
0.5 and 0.7, and a difference with the original values of up to 1.6 sigma.
\item $\beta$ as well as the latitudes of the fields studied here being small,
the line of sight
is almost parallel to the plane, which implies a slight anti-correlation
between $x_0$ and $\phi$. Furthermore, $\phi$ presents a slight correlation
with the density at the center $\rho_0$.
Moreover, the precision of $\phi$ depends on its value: it is
good when $\phi=14^\circ$, less good when $\phi$ is small (8.5$^\circ$)
and much worse when $\phi$ is high (37$^\circ$).
\item The cut-off radius $R_c$ is badly constrained, with
dispersions varying between 0.3 and 0.85 kpc.
\end{itemize}

%__________________________________________________________________

\section{Results of fits}

\subsection{First round}

The first round of fits presents a great degeneracy:
the angle $\phi$ is found to converge towards two distinct solutions.
While many fits converge to a median value in the range
5$^\circ$, 12.5$^\circ$, many others give a value very close to
0$^\circ$, which is inevitably biased because of the strict limit at
0$^\circ$ for this parameter. Medians and dispersions of the 2 groups
of solutions are tabled. The cut between the two partitions has
been fixed at $\phi$=3.5$^\circ$.

The degeneracy does not depend on the tested density profile, but is more
pronounced with Padova luminosity functions (especially
for age 12.6 Gyr) than with Bruzual \& Charlot LFs.

\subsubsection{Identification of badly fitted windows}

\begin{figure*}
\centering
\caption{Map of the mean $\chi_r$ (in number of $\sigma$)
by window over the best sets of parameters
obtained in the first round of fits using the Pad10 luminosity
function. The maps using the other LF are almost identical.}
\label{carte1}
\includegraphics[angle=-90,width=13cm]{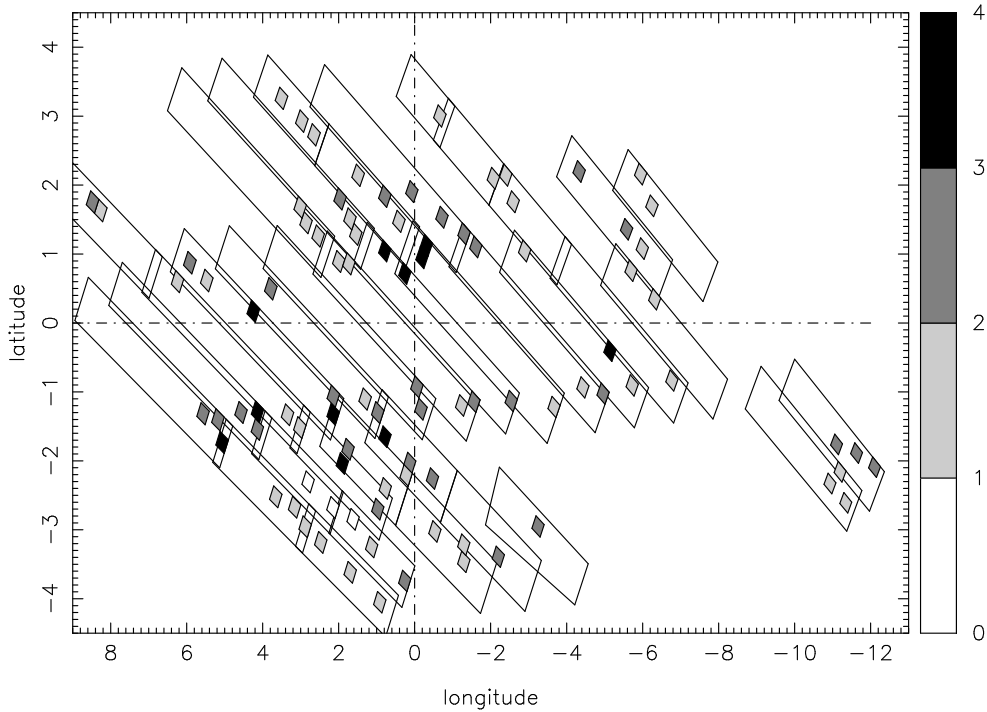}
\end{figure*}

Fig. \ref{carte1} presents the map of the mean $\chi_r$
(square root of $\chi^2$ per bin ; see appendix \ref{chir} for more
explanations) by window associated to the best parameters for the
Padova luminosity function at 10 Gyr.
The maps related to the different LF are similar,
and the badly fitted windows are mostly the same, even if the $\chi_r$ values
are a little smaller for the Bruzual \& Charlot luminosity functions.
Mean $\chi_r$ are calculated over all the fits without taking into account
the degeneracy.

Two groups of bad fields can be extracted:
\begin{itemize}
\item The first group of 8 windows is located at positive longitudes
(3$^\circ\leq$l$\leq$6$^\circ$) and negative latitudes
(-2$^\circ\leq$b$\leq$0.5$^\circ$), which contributes around 25\%
of the global likelihood. All these fields present a significant excess
compared to model counts.
A probable explanation is the following: these windows are amongst
the most extincted of all the studied fields. Firstly, this implies that
a poor modeling of the extinction distribution, fitted using
the Bruzual \& Charlot BC10 luminosity function, has a larger effect on
the star counts than expected.
Secondly, because of this high extinction,
the observed stars are intrinsically brighter
than in other directions, and the poor agreement may be caused by a
problem in the model for this range of absolute magnitudes, which has
almost no effect on the star counts in other locations.

\item A second group of 10 windows, located around the galactic
center, contributes around 25\% of the global likelihood. All these
fields present a large deficiency in the star count compared to the
model. This must be due to the
absence of a stellar population in the Besan\c{c}on model of the Galaxy,
the inner bulge, distinct from the outer bulge that we are studying here,
and confined to the first 1 or 2 degrees around the Galactic center.
The presence of such a central population is mentioned by
Ibata \& Gilmore (1995) and Frogel et al. (1999). The nuclear bar
evoked by Alard (2001) could correspond to the same population.
\end{itemize}

These two different groups of badly fitted windows represent less than 20\% of
the total number of fields but contribute more than half of the
global likelihood. Furthermore, the degeneracy in $\phi$ is mostly implied by
them.
As the aim of this study is to obtain a large scale description of the
outer bulge and thin disc populations, we prefer to remove these fields
and make new fits without them.

\subsection{Second round}

Tables \ref{table32} to \ref{table30}
show the best parameters related to the second round fits
for the 5 luminosity functions. For a given LF and a given density
profile, the medians and dispersions around these medians of the 20$\times m$
best values of parameters have been calculated.

One can see that the likelihood has decreased strongly.
The degeneracy in $\phi$
has almost disappeared with the Bruzual \& Charlot BC10 and BC12 and Padova
Pad7.9 functions, and with the Padova Pad10 LF when the S
density profile is used. It is less important but still present for the
Padova Pad10 (using E and G luminosity functions) and Pad12.6 functions.
As the fits converge away from 0$^\circ$, the limit between
the two groups of the degeneracy has been moved from $\phi$=3.5$^\circ$
to $\phi$=4.5$^\circ$.

\begin{table*}
\centering
\caption{\small
Results of the second round of fits using the Pad7.9 luminosity function.
The table is composed of 3 subtables, one for each density profile
($E$ $G$ and $S$). Each subtable is formed by two couples of lines
giving the median ($\mu$) and dispersion ($\sigma$) of the values.
Each pair of lines corresponds to a group of solutions (see text):
fits with $\phi\geq$4.5$^\circ$ (upper) and fits with $\phi <$4.5$^\circ$
(lower). The number of fits of each group is given at the beginning
of the first line. $L$ and $\chi_r$ mean global log-reduced
likelihood and square root of $\chi^2$ per bin, respectively (see Appendix
\ref{vrais} for more explanations).}
\label{table32}
\begin{tabular}{cccccccccccccccc}
\hline
\hline
\multicolumn{3}{c}{pad7.9} & $\phi$ & $\beta$ & $x_0$ & $y_0$ & $z_0$ &
$\rho_0$ & $R_c$ & $R_d$ & $R_h$ & C$_\parallel$ & C$_\perp$
& $L$ & $\chi_r$\\
\multicolumn{3}{c}{ } & $^\circ$ & $^\circ$ & kpc & kpc & kpc &
$\star$.pc$^{-3}$ & kpc & kpc & kpc & & & & \\
\hline
& 16 & $\mu_i$ & 7.1 & 0.6 & 1.35 & 0.41 & 0.32 & 18.62 & 3.66 & 2.44 & 1.25
& 2.969 & 2.804 & -1799 & 1.79\\
E & & $\sigma_i$ & 2.4 & 1.1 & 0.25 & 0.08 & 0.02 & 2.42 & 1.04 & 0.10 & 0.11
& 0.699 & 1.146 & 43 & 0.02\\
& 4 & $\mu_i$ & 2.6 & 0.0 & 1.73 & 0.56 & 0.32 & 16.56 & 2.23 & 2.46 & 1.24
& 3.392 & 1.112 & -1769 & 1.77\\
 & & $\sigma_i$ & 0.3 & 0.5 & 0.09 & 0.03 & 0.02 & 0.48 & 0.23 & 0.14 & 0.13
& 0.630 & 0.090 & 20 & 0.01\\
\hline
& 17 & $\mu_i$ & 10.6 & 0.4 & 1.60 & 0.47 & 0.40 & 9.84 & 3.44 & 2.30 & 1.27
& 3.201 & 3.751 & -1872 & 1.82\\
G & & $\sigma_i$ & 3.8 & 0.9 & 0.18 & 0.06 & 0.05 & 1.55 & 0.53 & 0.13 & 0.16
& 0.759 & 1.085 & 69 & 0.04\\
& 3 & $\mu_i$ & 2.0 & 0.3 & 2.16 & 0.78 & 0.39 & 8.78 & 2.78 & 2.46 & 1.31
& 2.675 & 1.021 & -1752 & 1.76\\
& & $\sigma_i$ & 0.3 & 0.9 & 0.39 & 0.10 & 0.01 & 1.31 & 0.20 & 0.04 & 0.17
& 0.914 & 0.250 & 111 & 0.05\\
\hline
 & 20 & $\mu_i$ & 10.6 & 0.8 & 1.82 & 0.53 & 0.45 & 11.48 & 3.71 & 2.35
& 1.31 & 3.375 & 3.489 & -1790 & 1.79\\
S & & $\sigma_i$ & 3.0 & 0.9 & 0.17 & 0.06 & 0.02 & 0.73 & 0.71 & 0.09
& 0.09 & 0.659 & 1.028 & 19 & 0.01\\
& 0 & \multicolumn{14}{c}{ }\\ 
\hline
\end{tabular}
\end{table*}

\begin{table*}
\centering
\caption{\small
Results of the second round of fits using the Pad10 luminosity function.}
\label{table34}
\begin{tabular}{cccccccccccccccc}
\hline
\hline
\multicolumn{3}{c}{pad10} & $\phi$ & $\beta$ & $x_0$ & $y_0$ & $z_0$ &
$\rho_0$ & $R_c$ & $R_d$ & $R_h$ & C$_\parallel$ & C$_\perp$
& $L$ & $\chi_r$\\
\multicolumn{3}{c}{ } & $^\circ$ & $^\circ$ & kpc & kpc & kpc &
$\star$.pc$^{-3}$ & kpc & kpc & kpc & & & & \\
\hline
& 9 & $\mu_i$ & 5.4 & 0.0 & 1.54 & 0.42 & 0.32 & 18.92 & 3.35 & 2.33 & 1.30
& 2.833 & 2.112 & -1921 & 1.85\\
E & & $\sigma_i$ & 1.9 & 0.5 & 0.11 & 0.04 & 0.06 & 1.50 & 0.73 & 0.14 & 0.18
& 0.788 & 1.514 & 54 & 0.03\\
& 11 & $\mu_i$ & 2.6 & 0.5 & 1.79 & 0.54 & 0.31 & 18.78 & 2.54 & 2.38 & 1.30
& 3.016 & 1.044 & -1862 & 1.82\\
& & $\sigma_i$ & 0.3 & 0.7 & 0.11 & 0.05 & 0.02 & 1.25 & 0.34 & 0.11 & 0.17
& 0.776 & 0.212 & 30 & 0.02\\
\hline
 & 8 & $\mu_i$ & 5.4 & 0.2 & 1.93 & 0.51 & 0.39 & 9.69 & 3.68 & 2.33 & 1.18
& 2.640 & 2.451 & -1957 & 1.87\\
G & & $\sigma_i$ & 2.4 & 0.5 & 0.22 & 0.05 & 0.02 & 0.93 & 0.84 & 0.18 & 0.20
& 0.689 & 1.022 & 61 & 0.03\\
& 12 & $\mu_i$ & 2.0 & 0.2 & 2.19 & 0.62 & 0.39 & 9.22 & 2.76 & 2.51 & 1.09
& 2.666 & 1.373 & -1886 & 1.83\\
& & $\sigma_i$ & 0.7 & 0.5 & 0.12 & 0.07 & 0.02 & 0.37 & 0.98 & 0.12 & 0.19
& 0.564 & 0.289 & 51 & 0.03\\
\hline
 & 17 & $\mu_i$ & 8.6 & 0.1 & 2.07 & 0.54 & 0.47 & 11.45 & 3.11 & 2.40 & 1.19
& 2.839 & 3.521 & -1934 & 1.86\\
S & & $\sigma_i$ & 2.0 & 0.6 & 0.18 & 0.04 & 0.02 & 0.77 & 0.74 & 0.10 & 0.12
& 0.836 & 0.803 & 33 & 0.01\\
& 3 & $\mu_i$ & 3.1 & -0.1 & 2.35 & 0.69 & 0.45 & 11.06 & 2.23 & 2.51 & 1.12
& 3.646 & 1.545 & -1900 & 1.83\\
& & $\sigma_i$ & 0.2 & 1.1 & 0.09 & 0.04 & 0.02 & 0.77 & 0.21 & 0.06 & 0.11
& 0.486 & 0.250 & 12 & 0.01\\
\hline
\end{tabular}
\end{table*}

\begin{table*}
\centering
\caption{\small
Results of the second round of fits using the Pad12.6 luminosity function.}
\label{table36}
\begin{tabular}{cccccccccccccccc}
\hline
\hline
\multicolumn{3}{c}{pad12.6} & $\phi$ & $\beta$ & $x_0$ & $y_0$ & $z_0$ &
$\rho_0$ & $R_c$ & $R_d$ & $R_h$ & C$_\parallel$ & C$_\perp$
& $L$ & $\chi_r$\\
\multicolumn{3}{c}{ } & $^\circ$ & $^\circ$ & kpc & kpc & kpc &
$\star$.pc$^{-3}$ & kpc & kpc & kpc & & & & \\
\hline
& 11 & $\mu_i$ & 4.5 & 0.3 & 1.74 & 0.41 & 0.33 & 18.01 & 3.39 & 2.42
& 1.13 & 2.380 & 1.959 & -2151 & 1.95\\
E & & $\sigma_i$ & 0.8 & 0.6 & 0.22 & 0.04 & 0.03 & 2.40 & 0.61 & 0.08
& 0.09 & 0.701 & 1.145 & 53 & 0.03\\
& 9 & $\mu_i$ & 2.5 & 0.4 & 1.92 & 0.46 & 0.31 & 18.32 & 2.86 & 2.48
& 1.22 & 2.597 & 1.375 & -2085 & 1.92\\
& & $\sigma_i$ & 0.5 & 0.3 & 0.20 & 0.04 & 0.02 & 1.72 & 0.39 & 0.14
& 0.19 & 0.516 & 0.154 & 33 & 0.02\\
\hline
 & 11 & $\mu_i$ & 6.9 & -0.5 & 2.07 & 0.49 & 0.41 & 9.97 & 3.27 & 2.37
& 1.17 & 2.570 & 3.299 & -2200 & 1.97\\
G & & $\sigma_i$ & 2.8 & 0.8 & 0.29 & 0.05 & 0.07 & 1.72 & 0.44 & 0.10
& 0.13 & 0.783 & 0.927 & 115 & 0.05\\
 & 9 & $\mu_i$ & 1.8 & 0.1 & 2.36 & 0.64 & 0.39 & 9.27 & 2.44 & 2.55
& 1.02 & 3.079 & 1.204 & -2106 & 1.93\\
& & $\sigma_i$ & 0.5 & 1.0 & 0.13 & 0.05 & 0.03 & 0.61 & 0.61 & 0.09
& 0.13 & 0.783 & 0.295 & 51 & 0.03\\
\hline
 & 10 & $\mu_i$ & 7.9 & -0.4 & 2.22 & 0.53 & 0.45 & 11.39 & 3.15 & 2.39
& 1.16 & 3.367 & 3.618 & -2181 & 1.97\\
S & & $\sigma_i$ & 2.0 & 0.9 & 0.21 & 0.04 & 0.02 & 0.41 & 0.66 & 0.06 
& 0.08 & 0.646 & 1.115 & 40 & 0.02\\
& 10 & $\mu_i$ & 2.6 & 0.4 & 2.61 & 0.67 & 0.46 & 10.76 & 3.04 & 2.41
& 1.16 & 3.082 & 1.426 & -2106 & 1.93\\
& & $\sigma_i$ & 0.5 & 1.1 & 0.21 & 0.07 & 0.03 & 0.58 & 0.82 & 0.08
& 0.10 & 0.533 & 0.240 & 54 & 0.02\\
\hline
\end{tabular}
\end{table*}

\begin{table*}
\centering
\caption{\small
Results of the second round of fits using the BC10 luminosity function.}
\label{table38}
\begin{tabular}{cccccccccccccccc}
\hline
\hline
\multicolumn{3}{c}{BC10} & $\phi$ & $\beta$ & $x_0$ & $y_0$ & $z_0$ &
$\rho_0$ & $R_c$ & $R_d$ & $R_h$ & C$_\parallel$ & C$_\perp$
& $L$ & $\chi_r$\\
\multicolumn{3}{c}{ } & $^\circ$ & $^\circ$ & kpc & kpc & kpc &
$\star$.pc$^{-3}$ & kpc & kpc & kpc & & & & \\
\hline
& 17 & $\mu_i$ & 8.5 & 0.7 & 1.26 & 0.46 & 0.35 & 21.39 & 2.38 & 2.45 & 1.28
& 3.235 & 3.715 & -2048 & 1.92\\
E & & $\sigma_i$ & 2.3 & 1.3 & 0.14 & 0.06 & 0.03 & 1.63 & 0.36 & 0.11 & 0.13
& 0.771 & 1.206 & 26 & 0.01\\
& 3 & $\mu_i$ & 3.3 & -0.2 & 1.31 & 0.58 & 0.33 & 22.19 & 2.38 & 2.47 & 1.27
& 3.424 & 1.348 & -2031 & 1.92\\
& & $\sigma_i$ & 0.3 & 0.7 & 0.18 & 0.08 & 0.01 & 1.63 & 0.42 & 0.19 & 0.28
& 0.331 & 0.228 & 18 & 0.01\\
\hline
& 20 & $\mu_i$ & 12.1 & 1.3 & 1.36 & 0.50 & 0.39 & 12.93 & 3.56 & 2.47 & 1.25
& 3.340 & 3.817 & -2102 & 1.94\\
G & & $\sigma_i$ & 1.7 & 1.2 & 0.10 & 0.02 & 0.02 & 0.82 & 0.78 & 0.12 & 0.15
& 0.762 & 0.351 & 21 & 0.01\\
& 0 & \multicolumn{14}{c}{ }\\ 
\hline
& 20 & $\mu_i$ & 12.4 & 1.6 & 1.55 & 0.57 & 0.44 & 14.91 & 4.00 & 2.52 & 1.27
& 3.801 & 4.149 & -2027 & 1.92\\
S & & $\sigma_i$ & 1.5 & 0.7 & 0.06 & 0.03 & 0.01 & 0.48 & 0.63 & 0.10 & 0.14
& 0.423 & 0.704 & 14 & 0.01\\
& 0 & \multicolumn{14}{c}{ }\\
\hline
\end{tabular}
\end{table*}

\begin{table*}
\centering
\caption{\small
Results of the second round of fits using the BC12 luminosity function.}
\label{table30}
\begin{tabular}{cccccccccccccccc}
\hline
\hline
\multicolumn{3}{c}{BC12} & $\phi$ & $\beta$ & $x_0$ & $y_0$ & $z_0$ &
$\rho_0$ & $R_c$ & $R_d$ & $R_h$ & C$_\parallel$ & C$_\perp$
& $L$ & $\chi_r$\\
\multicolumn{3}{c}{ } & $^\circ$ & $^\circ$ & kpc & kpc & kpc &
$\star$.pc$^{-3}$ & kpc & kpc & kpc & & & & \\
\hline
& 18 & $\mu_i$ & 8.0 & -0.8 & 1.40 & 0.46 & 0.36 & 21.56 & 2.24 & 2.44 & 1.22
& 3.607 & 3.257 & -1978 & 1.88\\
E & & $\sigma_i$ & 1.2 & 1.8 & 0.13 & 0.04 & 0.03 & 1.84 & 0.42 & 0.12 & 0.16
& 0.594 & 0.666 & 36 & 0.02\\
& 2 & $\mu_i$ & 3.5 & 1.1 & 1.49 & 0.61 & 0.33 & 22.80 & 1.96 & 2.34 & 1.33
& 4.168 & 1.173 & -1936 & 1.87\\
& & $\sigma_i$ & 0.2 & 0.3 & 0.03 & 0.02 & 0.00 & 0.25 & 0.07 & 0.03 & 0.03
& 0.072 & 0.069 & 1 & 0.00\\
\hline
& 19 & $\mu_i$ & 11.8 & 0.6 & 1.41 & 0.48 & 0.37 & 14.26 & 3.49 & 2.37 & 1.28
& 4.080 & 4.163 & -1954 & 1.87\\
G & & $\sigma_i$ & 2.3 & 0.7 & 0.08 & 0.04 & 0.02 & 0.66 & 0.85 & 0.11 & 0.11
& 0.741 & 0.896 & 22 & 0.01\\
& 1 & $\mu_i$ & 2.7 & 0.5 & 1.61 & 0.73 & 0.38 & 14.09 & 1.99 & 2.31 & 1.41
& 3.352 & 1.154 & -1958 & 1.88\\
& & $\sigma_i$ & 0.1 & 0.1 & 0.01 & 0.01 & 0.01 & 0.12 & 0.03 & 0.01 & 0.01
& 0.022 & 0.015 & 1 & 0.00\\
\hline
& 20 & $\mu_i$ & 12.3 & 0.9 & 1.58 & 0.55 & 0.45 & 16.79 & 4.01 & 2.35 & 1.36
& 3.654 & 4.106 & -1908 & 1.86\\
S & & $\sigma_i$ & 1.2 & 0.5 & 0.06 & 0.02 & 0.01 & 0.62 & 0.51 & 0.12 & 0.14
& 0.496 & 0.581 & 19 & 0.01\\
& 0 & \multicolumn{14}{c}{ }\\
\hline
\end{tabular}
\end{table*}

The second group of fits (with $\phi <$4.5$^\circ$), which are much less
numerous in the second round,
is now considered an artefact,
and only the other group of fits will be taken into account.

\subsection{Density profiles and luminosity functions}

Some conclusions can be deduced from Tables \ref{table32} to \ref{table30}:
\begin{itemize}
\item \emph{Density profiles}:
the gaussian function ($G$) (the least peaked at the center)
is the worst of
the three bulge density profiles tested. $E$ and $S$ functions reach similar
conclusions
but give slightly different best values, especially for $\phi$ and its
correlated parameters. The best agreement is obtained with the $S$ profile.

\item \emph{Luminosity functions}:
At similar ages, the Bruzual \& Charlot luminosity functions
give better agreement than the Padova (Girardi et al. 2002) ones. However,
the best LF used come from Padova models with
a bulge age of 7.9 Gyr.
The Bruzual \& Charlot isochrone at a similar age was not available.
This is the youngest tested value, which may mean
that the bulge age might be smaller.

\item \emph{Influence of the LF and density profile on best parameters}:
the thin disc scale length $R_d$ and $R_h$ determinations are robust
over the different luminosity functions and density
profiles which have been tested. This is not the same for the
bulge parameters. For instance, the orientation angle $\phi$ is different
with the $E$ profile and with the two others, and shows a significant
dependency
on the luminosity function: it is larger when the Bruzual \& Charlot
LFs are used, and decreases when the age increases. Correlated
parameters such as $R_c$ or $x_0$ are also affected.
\end{itemize}

\subsection{Quality of fits and correlations}

\begin{figure*}
\centering
\caption{Map of the mean $\chi_r$ by window
related to the Pad10 luminosity function.}
\label{carte3}
\includegraphics[angle=-90,width=13cm]{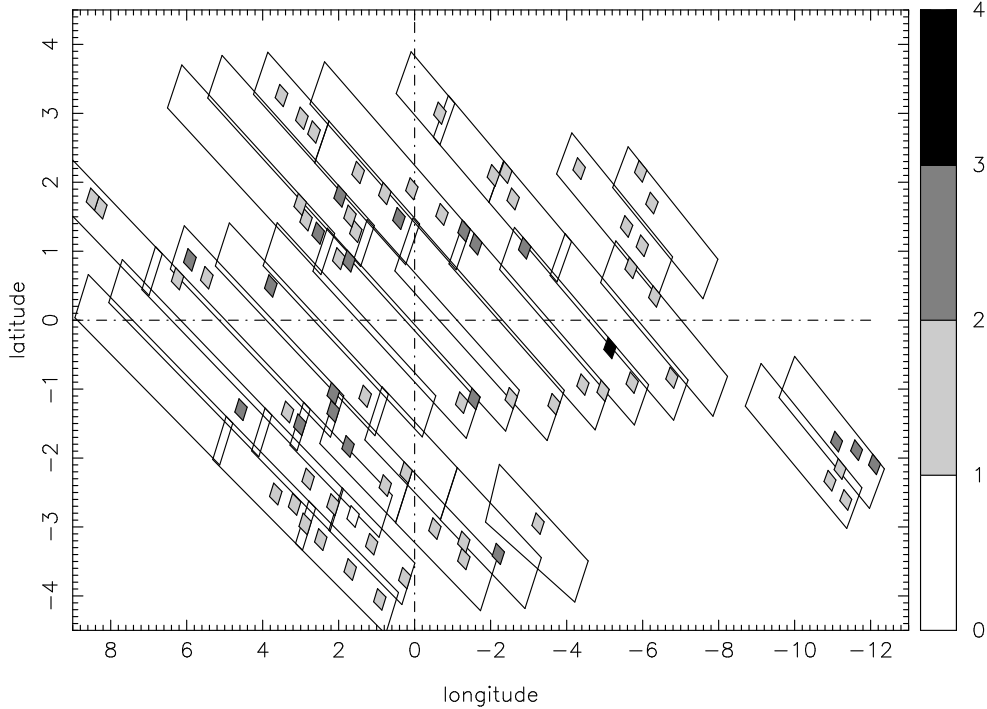}
\end{figure*}

Fig. \ref{carte3} shows the map of $\chi_r$ (square root of the $\chi^2$
per bin) associated to the fits using the Padova luminosity functions
at 10 Gyr. The other maps related to other LFs are similar.
One can see that the agreement is good (less than 3 sigmas) for all but
one or two windows.

\subsubsection{Accuracy of the results}

The dispersions obtained for the majority of parameters are small,
however some of them are not well constrained,
such as the bulge shape coefficients
$C_\perp$ and $C_\parallel$ and the cut-off radius $R_c$.
Nevertheless, this is not a serious problem. A small
variation of these parameters does not change significantly the bulge spatial
distribution: $C_\parallel$ and $C_\perp$ give only a general indication of
the shape of the outer bulge, and do not have significant influence on other
parameters, except a little on the bulge minor scale lengths $y_0$ and $z_0$
with which they are correlated. As for the cut-off radius, the scale length
$x_0$ being short, the density varies by 10\% to 20\% at the distance $R_c$,
and the effect of the cut-off is not very strong.

\subsubsection{Correlations}

Table \ref{correlations} shows the matrix of correlations around the
best parameters for the second round of fits (group with $\phi >$ 4.5$^\circ$
only). The choice of the used luminosity function and density profile
does not have any significant influence on the correlations, the matrix
corresponds to the mean values over all the (LF,profile) pairs.
For a given LF and a given density profile, correlations are computed, using
all the fits, as follows: let ($\xi^k$) be the $m\times 20$ best points
and $w_k$ their weight (deduced from their likelihood) ; let us take two axis
$i$ and $j$, the coordinates $\xi_i$ and $\xi_j$ of the points on these
two axes, their weighted means $m_i=\sum_k w_k\cdot\xi^k_i$ and
$m_j=\sum_k w_k\cdot\xi^k_j$, and their
weighted dispersions $s_i=\sum_k w_k\cdot(\xi^k_i-m_i)^2$ and
$s_j=\sum_k w_k\cdot(\xi^k_j-m_j)^2$ ;
then the correlation between the parameters $i$ and $j$ is equal to
$\frac{\sum_k w_k\cdot(\xi^k_i-m_i)^2><(\xi^k_j-m_j)^2}{s_i s_j}$.

\begin{table*}
\centering
\caption{\small
Mean values over the 5 luminosity functions and the 3 density profiles
of the correlations of the first group fits of the second round.
Values over $\pm$0.7 are written in boldface.}
\label{correlations}
\begin{tabular}{ccccccccccc}
\hline
\hline
& $\beta$ & $x_0$ & $y_0$ & $z_0$ &
$\rho_0$ & $R_c$ & $R_d$ & $R_h$ & C$_\parallel$ & C$_\perp$\\
\hline
$\phi$ & 0.0 & \textbf{-0.7} & \textbf{-0.7} & -0.3 & 0.6 & 0.2 & -0.2 & 0.0
& 0.5 & 0.6\\
$\beta$ & & 0.1 & -0.1 & -0.4 & 0.0 & 0.1 & -0.1 & 0.3 & 0.1 & -0.1\\
$x_0$ & & & 0.6 & 0.3 & \textbf{-0.9} & -0.4 & 0.1 & 0.1 & -0.4 & -0.2\\
$y_0$ & & & & 0.6 & -0.5 & -0.5 & 0.2 & 0.0 & -0.5 & -0.5\\
$z_0$ & & & & & -0.3 & -0.3 & 0.2 & -0.2 & -0.6 & -0.2\\
$\rho_0$ & & & & & & 0.3 & -0.1 & 0.0 & 0.2 & 0.1\\
$R_c$ & & & & & & & -0.1 & 0.1 & 0.1 & 0.1\\
$R_d$ & & & & & & & & \textbf{-0.8} & -0.3 & -0.2\\
$R_h$ & & & & & & & & & 0.0 & 0.1\\
$C_\parallel$ & & & & & & & & & & 0.3\\
\hline
\end{tabular}
\end{table*}

The main conclusions given by these correlations are the following:
\begin{itemize}
\item The disc scale lengths $R_d$ and $R_h$
are not correlated with bulge parameters but are strongly anticorrelated:
the variation in the
density due to decreasing one of the two parameters can be compensated
by increasing the other one.
This results in an incertainty of the hole scale length $R_h$,
though without questioning the existence of the hole.
\item As we have already shown with the tests (see Sect. \ref{qualite}),
the orientation angle $\phi$ is anti-correlated with parameters
such as $x_0$ and $y_0$.
\item The bulge major scale length $x_0$ and central density $\rho_0$,
whose variations have opposite effects on the number of stars
at a given distance, are also strongly anti-correlated.
\end{itemize}

\subsection{Best parameters}\label{totaleffectif}

\begin{table*}
\centering
\caption{\small
Best values obtained from the second round of fits.
The first double line corresponds to the model giving the best agreement
with data, i.e. with the $S$ density profile and the Pad7.9 luminosity
function. The values of the second one are the mean and dispersions
of the best set of parameters obtained with each pair density profile /
luminosity function. Meaning of the parameters:
$\phi$ gives the orientation of the bulge major axis with respect to the
Sun - center direction ; $\beta$ is the angle between this bulge major axis
and the Galactic plane ; the bulge major scale length $x_0$, whose
significance depends on the density profile, has been replaced by $\hat{x}_0$,
which is the distance on the major axis at which the density is equal to
38.6\% of the central one ; $r_y=\frac{y_0}{x_0}$ and $r_z=\frac{z_0}{x_0}$
are respectively the axis ratios associated to the bulge minor scale lengths
$y_0$ and $z_0$ ; $N_{\mbox{tot}}$ gives the total number of
bulge stars ; $R_C$ is the cut-off radius of the outer bulge,
and $C_\parallel$ and $C_\perp$ correspond to its shape coefficients ;
$R_d$ and $R_h$ are respectively the scale lengths of the
thin disc and of its central hole.}
\label{toptop}
\begin{tabular}{ccccccccccc|cc}
\hline
\hline
& & $\phi$ & $\beta$ & $\hat{x}_0$ & $r_y$ & $r_z$ &
$N_{\mbox{tot}}$ & $R_c$ & C$_\parallel$ & C$_\perp$ & $R_d$ & $R_h$\\
& & $^\circ$ & $^\circ$ & kpc & & & $10^{10}\star$ & kpc & & & kpc & kpc\\
\hline
best & $\mu$ & 10.6 & 0.8 & 1.97 & 0.30 & 0.25 & 6.39 & 3.71 &
3.38 & 3.49 & 2.35 & 1.31\\
model & $\sigma$ & 3.0 & 0.9 & 0.17 & 0.02 & 0.01 & 1.9 & 0.71 &
0.09 & 0.09 & 0.67 & 1.03\\
\hline
mean best & $\mu$ & 9.4 & 0.6 & 1.74 & 0.31 & 0.26 & 8.24 & 3.25 &
3.40 & 3.68 & 2.40 & 1.26\\
models & $\sigma$ & 2.8 & 0.5 & 0.24 & 0.04 & 0.03 & 1.35 & 0.64
& 0.40 & 0.47 & 0.06 & 0.06\\
\hline
\end{tabular}
\end{table*}

Table \ref{toptop} gives values of the fitted parameters for
the best model (using the
Freudenreich (1998) sech$^2$ density profile and the Girardi et al. 2002
(Padova) luminosity function with an age of 7.9 Gyr), as well as the ones
obtained by calculating the mean of the best values over all the pairs of
density profile / luminosity function.
These two sets of parameters are consistent with each other.
We can therefore deduce that the configuration described
here is robust, and does not depend on the choice
of the LF and the density profile.

This configuration is the following:
\begin{itemize}
\item \emph{Outer bulge}: the outer bulge is prolate, with axis ratios
1 : 0.31$\pm$0.4 : 0.26$\pm$0.3 (values from mean best models),
and seems to be boxy in all directions
($C_\perp >$2, $C_\parallel >$2).
Its major axis lies almost in the Galactic plane
($\beta$ is consistent with the value 0$^\circ$) and is oriented about
10$^\circ$ with respect to the Sun - center direction.
It contains 82$\cdot 10^9\pm$20$\cdot 10^9$ stars,
which corresponds to a mass of
2.4$\pm$0.6$\cdot 10^{10} M_\odot$.
\item \emph{Thin disc}: the thin disc shows a large hole at its center.
With a disc scale length of about 2.4 kpc and a hole scale length of about
1.3 kpc, the in-plane density goes from zero at the center to its maximum
at a distance of 2 kpc, and then decreases until the cut-off at 14 kpc.
However, the behaviour of the disc in the inner 100 pc is not
constrained here.
\end{itemize}

%__________________________________________________________________

\section{Discussion}

Many studies have been made of the structure of the outer bulge region.
Here, we compare the description obtained by our fits with those
found in the literature.

\subsection{Disc hole}

The strong anti-correlation between the two disc parameters $R_d$ and $R_h$
forces us to be careful with the results, which may be biased.
Nevertheless, the hole radius value (assumed to be the radius of the
maximal disc density), $\approx$2 kpc, is large enough to be used as
evidence of the presence of a central hole in the inner thin disc.

This conclusion is in agreement with the studies of external galaxies
by Ohta et al. (1990) and Bagget et al. (1996),
who argued for the existence of holed or inner truncated discs
in most spiral barred galaxies,
as well as the Galactic disc models of Freudenreich (1998)
who studied the CORBE/DIRBE map and obtained a hole radius of 3 kpc,
L\'opez-Corredoira et al. (2001) who analysed DENIS data and found
a maximum disc density at about 2 kpc from the Galactic center,
and L\'epine \& Leroy (2000) who found that
a model including a central hole is in better agreement with
the COBE/DIRBE surface brightness distribution and the rotation curve.

However, a central hole or an inner truncation are not the only things
which can give a smaller disc density in the plane close to the center.
An alternative explanation is a vertical flare in the inner disc. 
The scale height of the disc may be higher there as an effect of the bar
potential. Therefore, assuming that the surface density of the disc is
constant, this flare leads to a decreasing of the disc density in and close to
the Galactic plane, as does a central hole.
For instance, L\'opez-Corredoira et al. (2004) found similar agreements
with their data in the plane in the range 2-8 kpc
using either a flaring disc or a holed one.
As a flaring disc model does not change the global density,
contrary to a holed 
disc model, a further study of the vertical distribution of the inner thin disc
might make it possible to settle between these two alternative models.

\subsection{Outer bulge age}

All the tested bulge luminosity functions have been built using the same
scheme: the bulge is composed of only one generation of stars, which are
rather old, and the mean metallicity has been assumed solar
but with a large dispersion of 0.5 dex.
These hypotheses are consistent with most constraints
found in the literature.

The best agreement in our fits is obtained with the luminosity function
from Girardi et al. (2002) with an age of 7.9 Gyr. This is the youngest tested
age. Therefore, we can assume that the bulge age is at least younger than
10 Gyr, and perhaps even younger than 8 Gyr,
which is in contradiction with the results of Zoccali et al. (2003) who
propose 10 Gyr as a minimum, but is in favour of Cole \& Weinberg (2002)
who claimed that the outer bulge is not older than 6 Gyr.
New luminosity functions with a younger age from
Bruzual \& Charlot (2003) as well as from Girardi et al. (2002) (Padova)
are now available and we shall attempt to use them in the future
to test the hypothesis of a younger age, or 
several mixed generations of stars, as well as the influence of
the assumed bulge metallicity.

\subsection{Outer bulge shape}

Our best density profiles are the sech$^2$ function from Freudenreich (1998)
and the exponential one proposed by Stanek et al. (1997), with a preference
for the sech$^2$ profile. On the contrary, the gaussian function,
the best model found by Dwek et al. (1995), gives the worst agreements.

The outer bulge as described by our best parameters is very boxy. This is
consistent with most other works on the subject, for instance
Weiland et al. (1994) and later articles
(Dwek et al. 1995, Freudenreich 1998 ...) from the same authors, which
studied integrated luminosities from the COBE/DIRBE near infrared map.
Moreover, the axis ratios, 1:0.30$\pm$0.02:0.25$\pm$0.01, which
show that the triaxial bulge is very prolate, are similar to values
obtained by Freudenreich (1998) (1:0.37:0.26), Dwek et al. (1995),
Bissantz \& Gerhard (2002) (1:0.3-0.4:0.3, also from the COBE/DIRBE map)
and Weiner \& Sellwood (1999) (1:0.33:0.33, using HI and CO data).
This description of the outer bulge as a boxy prolate spheroid makes it
similar to a bar.

Concerning its half-length, the cut-off radius $R_c$ is fitted to
3.71$\pm$0.71 kpc. Using $x_0$=1.82 kpc, assuming no cut-off,
the bulge density on the major axis is equal to 3\% of the central density
at 4.4 kpc ($R_c$+1$\sigma$) from the center, and 14\% at 3 kpc
($R_c$-1$\sigma$). This means that a cut-off at 3.71 kpc is too
far to be very pronounced. This explains the small precision in $R_c$.
The existence of such a cut-off may be related to the disc dynamics.
A cutoff may appear at the corotation associated with the spiral arms or
the molecular ring. However, the position found here is a bit too far out
and inaccurate to allow us to conclude that it is a dynamical cut-off.

\subsection{Outer bulge orientation}

As in most of the other related works, the angle $\beta$ found with our fits
is very small and consistent with 0$^\circ$, which means that the outer
bulge major axis almost lies in the Galactic plane.

On the contrary, our estimation of $\phi$ (i.e. the angle
between the bulge major axis and the Sun - center direction),
10.6$^\circ\pm$3$^\circ$, differs
from some previous works. These studies, based on
COBE/DIRBE surface brightness (Dwek et al. 1995,
Binney et al. 1997, Bissantz \& Gerhard 2002), kinematics studies (Feast \&
Whitelock 2000, Bissantz et al. 2002), or IRAS (Nikolaev \& Weinberg 1997,
Deguchi et al. 2002) and OGLE star counts analyses (Stanek et al. 1997),
give values around 20$^\circ$.
However our estimation is consistent with the 12$^\circ\pm$6$^\circ$
obtained by L\'opez-Corredoira et al. (2000), who analysed
star counts from near
infrared large scale survey as we did, and compatible with the
14$^\circ$ found by
Freudenreich (1998) and L\'epine \& Leroy(200), both using the COBE/DIRBE map,
or even the 16$^\circ\pm$2$^\circ$ obtained by Binney et al. (1991)
from gas kinematics.

It is not clear why different studies coming from the same data
(COBE/DIRBE) give somewhat different results, which vary between
$\phi=20^\circ-30^\circ$ and $\phi=14^\circ-16^\circ$.
The low sensitivity to bulge stars at negative longitudes,
because of their large distance from us, may explain the difficulty
in determining precisely the bulge orientation angle. Furthermore,
it should be noted that these studies were based on integrated flux
density, hence they were less sensitive to the distribution
of stars along the line of sight than the present star count analysis.

\section{Conclusion}

We constructed a Monte Carlo fitting method to determine the
spatial distribution of outer bulge and disc stars from comparisons between
DENIS K$_s$ and J-K$_s$ stars counts and simulations from the Besan\c{c}on
model of the Galaxy. Our best parameters show a thin disc that has a central
hole, as often seen in barred spiral galaxies,
and a boxy prolate outer bulge with a major axis lying almost in the
Galactic plane, as obtained by most of the other studies.
However, our orientation angle (with respect to the Sun - center direction),
$\phi=10.6^\circ\pm 3^\circ$, is slightly smaller than the mean value
found in the literature.

The best fit among the different bulge ages which have been tested is
the youngest: 7.9 Gyr.
Future fits involving deeper data and more fields close to the
Galactic plane are planned to test other luminosity functions,
with younger ages and varying metallicities.

This study of the inner Galaxy is based on near infrared data, using low
extinction windows.
This approach does not allow us to have fields very close to the Galactic plane
and center. The use of mid-infrared observations, like ISOGAL
(Omont et al. 2003) or GLIMPSE (Benjamin et al. 2003), or deep near
infrared surveys like WIRCAM at the CFHT, soon to be available, will allow us
to reach these fields. It would be interesting to be able to include these
fields, especially to study the population of the inner bulge.

The boxy prolate shape and a possible young bulge age are compatible with
a bar structure, and may be explained by a reduced time scale for stellar
formation. However this photometric study is not sufficient to
settle the nature of the formation scenario of the outer bulge.
Kinematical data would be needed to confirm the bar nature
of the outer bulge, and thence its scenario of formation.

%__________________________________________________________________

\appendix

\section{Semi-gaussian drawings}
\label{tiragesgaussiens}

At each iteration of the fitting program,
gaussian drawings determining the new $p$ points were made using different
values of dispersion at the right and at the left of the median point
on the eigenframe axes. The aim was to describe as fully as possible
the morphology of the maximum likelihood region and reproduce it in the new
drawings.

Let ($\xi^k$, $k$=1..$m$) be a family of $m$ points of the
normalized 11-D space of parameters.
These points are sorted with respect to their reduced log-likelihood
(see Appendix \ref{vrais}), $L_1$ and $L_m$ being respectively the
maximal and minimal likelihoods, and weighted, the weight $w_k$
being defined by:
$$w_k=\exp(-1.5\frac{L_k-L_1}{L_m-L_1})$$
This formula allows us to enhance (under a certain limit)
the contribution of best points in the following formul{\ae}.

Let $\overline{\xi}$ be the median of the $m$ weighted points and
$\overrightarrow{V_j}$ be the eigenvector $j$ of
their covariance matrix.
Their values on the $i$ coordinate in the initial frame are respectively
$\overline{\xi}_i$ and $V_{ij}$.
We call $\hat{\xi}^k_j$ the coordinate $j$ of $\xi^k$ in
the eigenframe ($\overline{\xi}$,\{$\overrightarrow{V}_j$,$j$=1..11\}).

\

We define the semi-dispersion $\sigma_-$ and $\sigma_+$ by:
\begin{center}
$\sigma_-=\sqrt{\sum_{\hat{\xi}_j^k<0} w_j\cdot(\hat{\xi}_j^k)^2}$ et
$\sigma_+=\sqrt{\sum_{\hat{\xi}_j^k\geq0} w_j\cdot(\hat{\xi}_j^k)^2}$
\end{center}

Let ($\zeta^l_j$,$l$=1..$p$,$j$=1..11) be a family of $p\times$11
coordinates obtained using a gaussian drawing of mean 0 and dispersion 1.
The coordinates ($\tilde{\xi}^l_i$,$l$=1..$p$) in the initial frame of
the $p$ new points to be tested are then deduced
from the following formula:
$$\tilde{\xi}^l_i=\overline{\xi}_i+
\sigma_-\sum_{\zeta^l_{j<0}} V_{ij} \zeta^l_i +
\sigma_+\sum_{\zeta^l_{j\geq 0}} V_{ij} \zeta^l_i$$

\section{Likelihood and $\chi^2$ formulae}\label{vrais}

	Generally, when the agreement between observations and an analytic
model is tested, only the data has Poisson noise, and usual likelihood or
$\chi^2$ formul{\ae} are used.
In the case of the Besan\c{c}on model, initial simulations
also have a Poisson noise. This particularity must be taken
into account in the formul{\ae}\ of likelihood and $\chi^2$, as well as
the fact that the model counts are deduced from initial simulations using
weightings.

	Let $i$ identify one of the bins. Correlations between adjacent bins
are neglected. Let $y_i$ represent data star counts and $z_i$ be model counts.
$y_i$ (integer) follows a Poisson law around $Y_i$ (unknown).
$z_i$ (real) is obtained from an initial simulation star count
$z_i=\alpha {z_0}_i$,
$\alpha$ being the \emph{weight}\footnote{The case $\alpha =0$ is
excluded.} and ${z_0}_i$ (integer) following a Poisson law around
${Z_0}_i=\frac{1}{\alpha} Z_i$ ($Z_i$ and ${Z_0}_i$ being real).

\subsection{Reduced log-likelihood}

	Hereafter, probabilities (for discrete variables) will be
noted $P$ and densities of probabilities (for continuous variables)
will be noted $f$.

	By definition, the likelihood $\mathcal{L}_i$
(Kendall \& Stuart 1973) is the probability for an
observed star count to take the value $y_i$, assuming that the model is
correct, i.e. $Y_i$=$Z_i$=$\alpha {Z_0}_i$. One has:

\begin{equation}\label{eq1}
\mathcal{L}_i=P(y_i/z_i)=\int P(y_i/Z_i) f(Z_i/z_i) dZ_i 
\end{equation}
\begin{description}
\item[] $y_i$ follows a Poisson law around $Z_i$, so:
\begin{equation}\label{eq2}
P(y_i/Z_i)=\frac{1}{y_i!} {Z_i}^{y_i} e^{-Z_i}
\end{equation}

\item[] One has: $f(Z_i/z_i)=\frac{1}{\alpha} f({Z_0}_i/{z_0}_i)$.
According to Bayes' Theorem:
\begin{equation}\label{eq3}
f({Z_0}_i/{z_0}_i)=\frac{P({z_0}_i/{Z_0}_i)f({Z_0}_i)}
{\int P({z_0}_i/{Z_0}_i)f({Z_0}_i) d{Z_0}_i}
\end{equation}
Having no information a priori on ${Z_0}_i$, one can say that all $f({Z_0}_i)$
are equal\footnote{Bayes' Postulate}\
(at least around ${z_0}_i$, where $P({z_0}_i/{Z_0}_i)$ is not
negligible), so one can simplify Eq. \ref{eq3}:
$$f({Z_0}_i/{z_0}_i)=\frac{P({z_0}_i/{Z_0}_i)}
{\int P({z_0}_i/{Z_0}_i) d{Z_0}_i}$$
$P({z_0}_i/{Z_0}_i)=\frac{1}{{z_0}_i!} {{Z_0}_i}^{{z_0}_i} e^{-{Z_0}_i}$,
$\int P({z_0}_i/{Z_0}_i) d{Z_0}_i=1$, so:
\begin{equation}\label{eq4}
f(Z_i/z_i)=\frac{1}{\alpha}
\frac{1}{{z_0}_i!} {{Z_0}_i}^{{z_0}_i} e^{-{Z_0}_i}
\end{equation}

\item[] On substituting for $P(y_i/Z_i)$ from Eq. \ref{eq2},
for $f(Z_i/z_i)$ from Eq. \ref{eq4}, putting these into
Eq. \ref{eq1} and replacing $Z_i$ by
$\alpha {Z_0}_i$, one finds (with $\zeta =(1+\alpha) {Z_o}_i$):

\begin{eqnarray*}
\mathcal{L}_i & = &
\int \frac{1}{y_i!} (\alpha {Z_0}_i)^{y_i} e^{-\alpha {Z_0}_i}\frac{1}{\alpha}
\frac{1}{{z_0}_i!} {{Z_0}_i}^{{z_0}_i} e^{-{Z_0}_i} d(\alpha {Z_0}_i)
\nonumber \\
 & = & \frac{1}{y_i!{z_0}_i!}
\frac{\alpha^{y_i}}{(1+\alpha)^{y_i+{z_0}_i+1}}
\int \zeta^{y_i+{z_0}_i} e^{-\zeta} d\zeta \\
 & = & \frac{(y_i+{z_0}_i)!}{y_i!{z_0}_i!}
\frac{\alpha^{y_i}}{(1+\alpha)^{y_i+{z_0}_i+1}} \\
 & = & \frac{(y_i+{z_0}_i)!}{y_i!{z_0}_i!} {z_0}_i^{{z_0}_i+1}
\frac{z_i^{y_i}}{(z_i+{z_0}_i)^{y_i+{z_0}_i+1}} \\
\end{eqnarray*}
\end{description}

Actually, we usually use the reduced log-likelihood, which is the
natural logarithm of the likelihood minus the likelihood computed
with $z_i$ replaced by $y_i$:

\begin{eqnarray*}
L_i & = & \ln \frac{(y_i+{z_0}_i)!}{y_i!{z_0}_i!} {z_0}_i^{{z_0}_i+1}
\frac{z_i^{y_i}}{(z_i+{z_0}_i)^{y_i+{z_0}_i+1}} \\
 & - & \ln \frac{(y_i+{z_0}_i)!}{y_i!{z_0}_i!} {z_0}_i^{{z_0}_i+1}
\frac{y_i^{y_i}}{(y_i+{z_0}_i)^{y_i+{z_0}_i+1}} \\
 & = & y_i\ln (\frac{z_i}{y_i})-(y_i+{z_0}_i+1)\ln
(\frac{{z_0}_i+z_i}{{z_0}_i+y_i})
\end{eqnarray*}

	Finally, by summing all the bins, one obtains:

$$L=\sum_i [y_i\ln (\frac{z_i}{y_i})-(y_i+{z_0}_i+1)\ln
(\frac{{z_0}_i+z_i}{{z_0}_i+y_i})]$$

\

The relative difference between the reduced log-likelihood obtained here
and the usual one (i.e. without model noise:
$L_0=\sum_i [y_i\ln (\frac{z_i}{y_i})-(z_i-y_i)$) varies between 8\% and
12\%, which is not negligible.

\subsection{$\chi_r$}\label{chir}

	$y_i$ is a Poisson random variable with an average and variance $Y_i$.
This variance is unknown and its only approximation available is the value
$y_i$.

	$z_i$ is a random variable with an average $\alpha {Z_0}_i$
and a variance
$\alpha^2 {Z_0}_i=\frac{Z_i^2}{{Z_0}_i}$. This variance is unknown and its
only approximation available is the value $\frac{z_i^2}{{z_0}_i}$.

	So, $y_i-z_i$ is a variable with a variance
$\sigma_i^2\approx y_i+\frac{z_i^2}{{z_0}_i}$.
One therefore obtains the following formula of normalized residuals:
$$r_i=\frac{y_i-z_i}{\sqrt{y_i+\frac{z_i^2}{{z_0}_i}}}$$

	At the end, by summing over all the
color magnitude bins of a window, one obtains the $\chi^2$ per
field (or group of fields):

$$\chi^2_f=\sum_i r^2_i
=\sum_i \frac{(y_i-z_i)^2}{y_i+\frac{z_i^2}{{z_0}_i}}$$

	However, we would rather use the square root of the $\chi^2$ per bin
$\chi_{rf}=\sqrt{\frac{\chi^2}{n}}$ (with $n$ = number of bins)
wich corresponds to the mean dispersion
needed for random variables centered on the $y_i$ and used as model counts
to obtain on average the same value of $\chi^2_f$.

\

By summing over all the fields or groups of fields ($n_f$=88 in total),
one obtains the global $\chi^2$ and the square root of the $\chi$:
$$\chi_r=\frac{1}{n_f}\sum_f \chi_{rf}^2$$

\subsection{Reduced log-likelihoods vs $\chi_r$}

If the likelihood has been preferred to $\chi_r$ 
to extract best sets of parameters in the
fitting method, the $\chi_r$ is a more practical tool than likelihood
to 'read' the quality of a fit, because it gives directly,
in number of sigma, the mean distance between model and data.
That is why the values of $\chi_r$ have been calculated as well as
likelihoods.

\begin{figure}[h]
\centering
\caption{Reduced log-likelihoods (ordinates)
versus square root of $\chi^2$ per bin.}
\includegraphics[angle=-90,width=8cm]{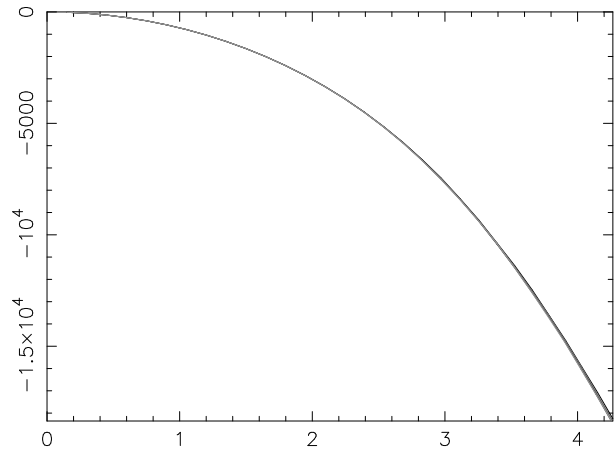}
\label{vrais-chi2}
\end{figure}

Fig. \ref{vrais-chi2} presents the values of
reduced log-likelihoods $L$ versus the square root of $\chi^2$ per bin $\chi_r$
obtained by random drawings.

\begin{acknowledgements}

	This study could not have been made whithout the previous work of
Emmanuel Chereul, when he was post-doctoral fellow at the Geneva Observatory.
The authors thank Guy Simon who was deeply involved in the
DENIS batch reduction process,
the whole DENIS staff and all people who observed and collected the
data. We thank Mathias Schultheis who gave us his extinction maps,
Edouard Oblak who helped us with the Monte Carlo fitting method, and
Doug Marshall who corrected the english. S\'ebastien Picaud
has benefited from an allowance from the R\'egion de Franche-Comt\'e.

\end{acknowledgements}

\end{document}